%file: D:\CUG4DM26.tex 

\centerline {\bf Fourth Generation fermions as candidates for Dark Matter}

\vskip5pt

\centerline  {D.J.Newman}

\vskip5pt

\centerline  {email: \it dougnewman276@gmail.com} 

\vskip20pt

\beginsection Abstract

Clifford Unification describes all the observed fundamental fermions
in terms of seven commuting elements of the $Cl_{7,7}$ Clifford algebra.
The eigenvalues of each commuting element define a binary quantum
number, which relates to a fermion property that is conserved in decays
and interactions. These include the quantum number
descriptions of a hitherto unobserved fourth generation G(4) of fermions,
which are predicted to have electric charges different to their observed G(1-3)
counterparts. This, together with quantum number conservation,
eliminates the possibility of interactions between G(4)
and G(1-3) fermions. Neutral G(4) composites are shown to
provide candidates for baryonic Dark Matter, which is identified
with the super-massive cores of galaxies. This could be
examined in the light of recent observations of 'Little Red Dots'
in the early Universe. Neutral leptonic G(4) composites provide
candidates for the Dark Matter component of galactic halos.
%\vfill\eject

\beginsection \S1. Introduction

There is considerable evidence
for the existence of very large amounts of
`Dark Matter' (DM) in galactic and inter-galactic space.
Until recently, DM has only been observed through its gravitational
effects, which show it to consist of electrically neutral
particles.  It has not yet been possible to
identify DM with composites of G(1-3) fermions,
or with any other of the predicted exotic particles. Many attempts,
based on sophisticated mathematical models, have been made
to identify other DM candidates, but experimental evidence
has yet to be obtained [1-4]. No previous attempts seem to 
have been made to identify DM with G(4) fermions.

Much effort has been made to
obtain the properties of DM particles
using earthbound experiments. Belyaev et al [3]
interpreted disappearing tracks at LHC as absorption of
particles by heavy DM, consistent with collisions with 
very heavy stable particles, but this interpretation
has yet to be confirmed.
Foot [4] has suggested that experimental data
available at that time could be explained in terms
of DM with two or more components. However, a 2021 
analysis of the COSINE-100 experiments [5] has thrown doubt
on this suggestion. Recent attempts,
for example [6,7,8], are based on the assumption of
specific properties that distinguish it from
ordinary matter. So far no statistically significant 
results have been obtained.

The existence of G(4) fermions is currently doubted by most particle
physicists because of the clear experimental evidence that
only three generations of neutrinos exist, e.g. see [9], p.438.
This has been interpreted as evidence that no
G(4) fermions exist. The more detailed analysis in [10] argues that
the Standard Model (SM) is inconsistent with the existence of G(4) fermions,
although some recent work does not accept this conclusion [11,12]. As these
discussions are not related to the G(4) fermions predicted in Appendix I,
their arguments are not relevant.

The starting point of this work is the theoretical prediction of
existence of G(4) fermions in Clifford Unification [13]. Details
of this prediction, given in this work, suggest
that G(4) fermions have properties associated with dark matter.

In this work                    
\item{\S2} justifies the predicted electric charges on G(4)
fermions given in Appendix I. \vskip2pt
\item{\S3} determines the properties of G(4) fermions \vskip2pt
\item{\S4} determines the structure of G(4) baryons\vskip2pt
\item{\S5} finds the structure of neutral G(4) composites, showing that
observed dark matter could take two forms, i.e. baryonic and leptonic \vskip2pt
\item{\S6} examines relationships between these composites and DM and provides
a possible explanation for the recent observations of `Little Red Dots'.\vskip2pt
\item{\S7} suggests the structure of galaxy cores and the production of
supermassive black holes \vskip2pt
\item{\S8} examines the cosmological significance of Clifford Unification\vskip2pt
\item{\S9} summarizes the results of this work and
its possible impact on the analysis and design of experiments.

\beginsection \S2. Prediction of the existence of, and charges on, G(4) fermions

Clifford Unification, based on the $Cl_{7,7}$ Clifford algebra, describes
the 4$\times$32 = 128 distinct states of the 32 fermions tabulated in
Appendix I. Only 4$\times$24 = 96 states in three generations of eight fermions
have been observed, confirming their calculated charges shown in this table.
The remaining fermions described in Appendix I are the 32 states of six quarks
and two leptons that form a fourth generation `G(4)' of fermions that has yet
to be observed. The predicted charges on the G(4) fermions given in Appendix I
look strange at first sight and require justification. Details of
their derivation are provided in Appendix II. An outline follows:

\item{[1]} The charges on all the G(1-3) observed fermions in can be expressed
in terms of a two component model giving the total charge as $\rm Q =Q_B+Q_C$,
where $\rm Q_B$ and $\rm Q_C$ are separately defined in terms of the fermion quantum
numbers. In particular, the charges on the G(1) electrons and neutrinos are
identified with $\rm Q_B = -1/2$ for all leptons, with $\rm Q_C = -1/2$ for
electrons and $\rm Q_C = +1/2$ for neutrinos. G(2) and G(3) have the same lepton
charges.
 
\item{[2]}  Appendix I shows the two leptons and six quarks in each of G(1-3)
are distinguished by the quantum numbers D=$\pm 1$, E$\pm 1$.
The quarks in all four generations have $\rm Q_B = 1/6$ and the
leptons have $\rm Q_B = -1/2$. It follows that the total $\rm Q_B = 1/6$
on all the fermions in any generation is zero. The
G(4) fermions are predicted to have the same lepton/quark structure as G(1-3)fermions,
ensuring the total number of G(4) fermions to be the same as that in each of G(1-3).

\item{[3]} The distinct G(1-3) and G(4) contributions to $\rm Q_C = 1/6$ fermion
charges are described by the F,G quantum numbers in the same way that the
quark/lepton contributions to $\rm Q_B$ are determined by the D,E quantum numbers,
(see Appendix II).

\item{[4]} Combining $\rm Q_B$ and $\rm Q_C$ gives the total charges on all
fermions in terms of their quantum numbers as
$$
\rm Q = Q_B + Q_C,\>\> where\>\> Q_B ={1\over 6}( D + E - BDE)\>\>
and\>\>Q_C = {1\over 2}( F + G - BFG )BC.                          \eqno(2.1)
$$ 
Inclusion of the quantum number B ensures that (2.1) is valid for both
anti-fermions and fermions.

\item{[5]} The table in Appendix I shows that no G(4) fermions have total charge
$\rm Q =Q_B+Q_C=0$, consistent with the experimental evidence [9,10] that no
G(4) neutrinos exist.

\item{[6]} Quarks have three colours and occupy only small regions of space-time.
The algebraic isomorphism between the lepton/quark distinction and the G(4)/G(1-3)
distinction suggests that G(1-3) fermions occupy relatively small regions of
space-time, whereas G(4) fermions are not subject to such a restriction.

\beginsection \S3. Properties of the predicted fourth generation fermions

The table in Appendix I includes eight currently unobserved types of G(4)
fermions, all of which are labelled by their charges:
two leptons l$(-2)$, l(1) and six quarks with additional colour labels, viz.
q$_b(-4/3)$, q$_r(-4/3)$, q$_g(-4/3)$ and q$_b(5/3)$, q$_r(5/3)$, q$_g(5/3)$.
The brackets give the predicted charges obtained from equation (2.1)
that gives the (experimentally) correct charges for all the
observed G(1-3) fermions and, in the case of G(4), are predicted by the analysis
in Appendix II.

It was shown in [13] that all the observed interactions
and decays of the G(1-3) fermions and mesons can be expressed in terms
of the conservation of seven binary quantum numbers. Further details, in
which fermion parities are taken into account, are given in [13].
Interactions between both G(4) and G(1-3) fermions, apart from gravity
and electromagnetic force, are impossible because of their different
relationships between fermion charges ($\rm Q_C$) and parity (C).

Nevertheless, there is one example of
quantum number conservation that involves only G(4) quarks and leptons.
In terms of the (CDEFG) quantum numbers given in Appendix I, this is:
$$
\{q_b(5/3):(1 1 1 \bar1 \bar1)\} + \{l(-2):(\bar1 \bar1 \bar1 \bar1 \bar1)\}
= \{q_b(-4/3):(\bar1 1 1 \bar1 \bar1)\} + \{l(1):((1 \bar1 \bar1 \bar1 \bar1))\},            \eqno(3.1) 
$$
with similar equations for r and g colour G(4) quarks. This interaction
could take place in either direction.

\beginsection \S4. G(4) baryons

Neutrons (n) and protons (p) have the G(1) quark structures n$\simeq$udd and
p$\simeq$uud, forming a doublet with the same electroweak interactions as the
electron/neutrino doublet. Analogous G(4) baryons b(2), b(-1) are
$$\eqalign{
	{\rm n(0)}=\>&{\rm u(2/3)d(-1/3)d(-1/3)\>\>\>} \leftrightarrow {\>\>\> b(-1)}
	 = { q(5/3)q(-4/3)q(-4/3)},\cr
	{\rm p(1)}=\>&{\rm u(2/3)u(2/3)d(-1/3)\>\>\>} \leftrightarrow {\>\>\> b(2)}
	= { q(-4/3)q(5/3)q(5/3)}}
\eqno (4.1)
$$
{\it showing that there are no neutral {\rm G(4)} baryons (constructed from three quarks)
analogous to the {\rm G(1)} neutrons.}

The algebraic description of strong interaction is the same for all four generations,
suggesting that a strong interaction (gluon) bonding between G(4) quarks is similar,
but possibly much stronger, than that between G(1) quarks. Greater masses might
produce significant gravitational bonding. Also, Compared to neutrons and protons,
the G(4) baryons ${ b(2)}$ and ${ b(-1)}$ defined above are subject to stronger
electrostatic bonding. Taking the mean inter-quark distance to be $r_0$, this can be
estimated as
$e^2[2(5/3)(4/3)-(4/3)^2]/r_0 =(8/3)e^2/r_0 $ for $b(2)$
and
$e^2[2(5/3)(4/3)-(5/3)^2]/r_0 =(5/3)e^2/r_0 $ for $b(-1)$.
This suggests that $b(2)$ is more stable than $b(-1)$, which is consistent
with the $q(-4/3)$ decay process identified in \S3.

\beginsection \S5. Neutral G(4) composites

G(1-3) matter is constructed from atoms, formed
from stable positively charged baryonic nuclei surrounded by clouds
of electrons. Similar neutral G(4) composites, constructed from baryons
and leptons are also possible, but two other types of neutral G(4)
composites, namely those formed entirely from either baryons or leptons,
are possible. Neutral baryonic composites would have structures of the
form $x(b(2)+2b(-1))$, where $x$ is any positive integer. Even values of
$x$, with a distorted tetrahedral structure and zero spin are of particular
interest. Similarly, neutral $x(l(-2)+2l(1))$ lepton composites,
again with even $x$, would also be possible.

Given the combination of gravitational, electrostatic and strong
local interactions, together with the small size and large mass of G(4)
baryons it is feasible that structures with large values of $x$ would
develop, providing candidates for the ultra-massive black holes with some
sort of crystalline structure forming
galaxy nuclei. Neutral lepton composites could provide the dark matter
observed in galaxy halos. However, as expected, all G(4) fermions were created in
equal numbers in the `big bang', a third type of neutral fermion composite
constructed from both baryons and leptons with the structures $x(b(2)+2l(-1))$ or
$x(l(2)+2b(-1))$, would also exist. These, like
atoms, would be subject to internal structural transitions, producing
radiation that would make them observable, rather than dark.

However, the numbers of these`atom-like' G(4) míxed baryon/lepton structures
could by now, have been reduced, to zero by the decay process given in
equation (3.1), i.e. $b(2) + l(-2) \to b(-1) + l(1)$.
If, over a very long period of time, one third of the original number
(say $y$) of all types of G(4) fermions interacted in this
way there would remain only $2y/3 b(2)$ and $4y/3 b(-1)$ baryons
and $2y/3 l(-2)$ and $4y/3 l(1)$ leptons, leaving no fermions
to construct `atom-like' nucleon/lepton neutral structures.
It is now possible, therefore, that all existing G(4) fermions form
neutral $x(b(2)+2b(-1))$ baryon composites and all leptons
are condensed into neutral $x(l(-2)+ 2l(1))$ lepton composites.

\beginsection \S6. Neutral G4 composites as observable Dark Matter

There remains, of course, the question why G(4) fermions have
never been observed, suggesting that they could be the
constituents of DM. In order to test this identification
it is necessary to relate the predicted properties of G(4) fermions
to the observed properties of DM. In particular it is necessary
to explain why G(4) matter is dark. G(1) fermions form stable atoms,
in which central nuclei composed of quarks has its positive
charge compensated by a surrounding cloud of electrons.
Decays of their excited states produce photons.
If neutral G(4) analogues of atoms existed,
they would almost certainly be observable.

Most of the experimental evidence for the existence of DM
relates to its distribution in the halos that surround the cores of
galaxies [14]. This could be attributed to neutral lepton composites.
It is generally agreed that galaxy cores are
black holes produced by high density DM, which can be attributed
to neutral baryon composites.
Recent observations of have provided information
on an early step in the formation of supermassive black hole cores [15].

It is generally accepted the black holes, especially the
super-massive black holes (SMBH)[16], with masses of more than
$10^6$ solar masses, form galaxy nuclei. The overall
DM to ordinary matter (OM) mass ratio, estimated
from halo observations, is
about $m_{DM}/m_{OM}= 5$. This suggests that dark matter
fermions to have approximately five times the
mass their G(1) counterparts, giving an estimate of
the mass of the $x=1$ neutral baryon composites as $3\times5 = 15$
proton masses. This could, however be an underestimate, because it
takes no account of the amount of dark matter in galaxy nuclei.
Composites with even values $x$ would be bosons with integral spins.
As there is no limit to $x$, they provide candidates for black holes and
produce baryonic galaxy nuclei of arbitrary size.
 
\beginsection \S7. Observations of G(4) composites

Galaxy halos [11] apparently contain considerable amounts of DM,
suggesting that this could be a space filling gas of $x=1$ G(4) neutral
lepton composites. G(1-3) would occupy only localised regions
in this gas. It also suggests a possible gravitational process for the
accretion of halo material in galaxy nuclei [17,18,19].

Galaxy nuclei that provide the gravitational field
that holds galaxies together, have been shown to be supermassive
black holes (SMBH). The mass at the centre of the Milky Way has been determined
to be 4.6$\times 10^6$ solar masses, and other galaxies have SMBH with as
much as $10^9$ solar masses. There is no possible way
that such massive galaxies could be constructed from G(1-3) matter.
On the other hand, G(4) baryonic composites, with the properties
described above, provide the perfect candidates for the formation of SMBH
by the accumulation of the primordial black holes, each of which were
produced by a very few G(4) baryons [20].

Recent observations of `Little Red Dots' in the early universe have
been explained in a number of ways. One explanation identifies them
as the early stage in the formation of supermassive black holes
by the accretion of ultra-strongly interacting dark
matter [14]. It is possible that the observed interaction is the
gravitational accretion of `atom-like' nucleon/lepton neutral G(4)
structures into a growing mass of baryonic matter.

\beginsection \S8.  Cosmological significance

It is generally accepted that matter was formed by the condensation of energy
in the `big bang'. It is usually associated with the production of equal numbers
of fermions and anti-fermions. This process is unlikely to have been instantaneous
over the whole of space and suggests that energy was converted into equal numbers
of fermions in specific regions, with cancellations in the charges on
different fermions formed in each region. The table in Appendix I show
cancellations in both $\rm \Sigma Q_B$ and$\rm \Sigma Q_C$ charges for
each of the four generations, making the total charge
$\rm \Sigma Q$ on the fermions in each of the four generations zero.
The $\rm Q_C$ cancellation is because the pairing of each fermion with
a fermion of opposite parity (C), which is likely to have occurred before
the lepton/quark split. There is no reason why created fermions
should be associated with a particular parity of the coordinate system
that has been constructed by human observers, so it is significant
that half of them are associated with each parity.
  
Relations between fermion charges in all four generations suggest
that the `big-bang' was a multi-step process in which the first step
was the separation of fermions and anti-fermions corresponding to the
direction of time and the quantum number B. The second step
was the distinction of positive and negative parities corresponding
to the quantum number C. Subsequent steps were the distinction between
leptons and quarks and the separation of G(1-3) and G(4) generations.
This separation also distinguishes time direction,
solving the problem of finding how the anti-fermions subsequently
`disappeared', and suggesting that fermions
and anti-fermions occupy different, i.e. future/past,
regions of space-time, corresponding to positive
and negative values of the quantum number B. This was followed
by the separation of fermion parities with opposite charges corresponding
to the positive and negative values of C. Later separations are
related to spatial divisions associated with the quark/lepton
distinction and the G(1-3)/G(4) distinction.

\beginsection \S9 Conclusions

The existence of a fourth generation of fermions is a
clear prediction of Clifford Unification [13]. This has led to
the identification of G(4) fermions with dark matter which
takes two forms: baryonic, forming galaxy cores
and leptonic, forming galaxy halos. This validates the conceptual
framework developed in [13], that shows fermion properties to be
determined by their interactions. This work
proposes that dark matter is composed of fourth generation
fermions. The electric charges on G(4) fermions have been
determined by employing the isomorphism of two $Cl_{2,2}$ sub-algebras
of $Cl_{7,7}$. These are very different to the charges on the three
observed generations, and are consistent with the existence of distinct
neutral G(4) composites that are composed entirely of baryons or leptons.
These properties have been shown to be consistent with the super-massive
black holes that form the cores of observed galaxies being formed
from neutral baryon composites and the dark
matter in galaxy haloes being formed from neutral lepton composites.

\vfill\eject

\beginsection  Appendix I: Charges and quantum numbers
 
$$\vbox
{\settabs 11 \columns
	\+&{\bf Quantum numbers, with B=1, A=$\pm 1$, for all four generations of fermions} \cr 
	\+&|||||||||||||||||||||||||||||||||||||||||\cr
	\+&quark&C&D&E&F&G$\>\>\>\>\>\>\>\rm gene$&$\rm ration$ & $\rm \bf Q_B$&$\rm \bf Q_C$&$\rm \bf Q$\cr
	\+&|||||||||||||||||||||||||||||||||||||||||\cr
	\+&${\rm u}_b$    &$-1$   &$\>\>1$&$\>\>1$&$\>\>1$ &$\>\>1$   &[1]&1/6&1/2&$\>\>2/3$\cr
	\+&${\rm u}_r$    &$-1$   &$\>\>1$&$-1$   &$\>\>1$ &$\>\>1$   &[1]&1/6&1/2&$\>\>2/3$\cr
	\+&${\rm u}_g$    &$-1$   &$-1$   &$\>\>1$&$\>\>1$ &$\>\>1$   &[1]&1/6&1/2&$\>\>2/3$\cr 
	\+ &&&\cr
	\+&${\rm d}_b$    &$\>\>1$&$\>\>1$&$\>\>1$&$\>\>1$ &$\>\>1$&$[1]$&1/6&$-1/2$&$-1/3$\cr
	\+& ${\rm d}_r$   &$\>\>1$&$\>\>1$&$-1$   &$\>\>1$ &$\>\>1$&$[1]$&1/6&$-1/2$&$-1/3$\cr
	\+&${\rm d}_g$    &$\>\>1$&$-1$   &$\>\>1$&$\>\>1$ &$\>\>1$&$[1]$&1/6&$-1/2$&$-1/3$\cr 
	\+&|||||||||||||||||||||||||||||||||||||||||\cr
	\+&${\rm c}_b $       &$-1$   &$\>\>1$&$\>\>1$&$\>\>1$    &$-1$   &[2]&1/6&$1/2$&$\>\>2/3$\cr
	\+&${\rm c}_r $       &$-1$   &$\>\>1$&$-1$   &$\>\>1$    &$-1$   &[2]&1/6&$1/2$&$\>\>2/3$\cr
	\+& ${\rm c}_g $      &$-1$   &$-1$  &$\>\>1$ &$\>\>1$    &$-1$   &[2]&1/6&$1/2$&$\>\>2/3$\cr 
	\+ &&&\cr
	\+& ${\rm s}_b $      &$\>\>1$&$\>\>1$&$\>\>1$&$\>\>1$    &$-1$   &$[2]$&1/6&$-1/2$& $-1/3$\cr
	\+&${\rm s}_r $       &$\>\>1$&$\>\>1$&$-1$   &$\>\>1$    &$-1$   &$[2]$&1/6&$-1/2$& $-1/3$\cr
	\+& ${\rm s}_g $      &$\>\>1$&$-1$   &$\>\>1$&$\>\>1$    &$-1$   &$[2]$&1/6&$-1/2$& $-1/3$\cr
	\+&|||||||||||||||||||||||||||||||||||||||||\cr		
	\+ &${\rm t}_b $  &$-1$   &$\>\>1$&$\>\>1$&$-1$  &$\>\>1$   &[3]&1/6&$1/2$&$\>\>2/3$\cr
	\+ &${\rm t}_r $  &$-1$   &$\>\>1$&$-1$   &$-1$  &$\>\>1$   &[3]&1/6&$1/2$&$\>\>2/3$\cr
	\+ &${\rm t}_g $  &$-1$   &$-1$   &$\>\>1$&$-1$  &$\>\>1$   &[3]&1/6&$1/2$&$\>\>2/3$\cr 
	\+ &&&\cr
	\+ &${\rm b}_b $  &$\>\>1$&$\>\>1$&$\>\>1$&$-1$  &$\>\>1$   &$[3]$&1/6&$-1/2$&$-1/3$\cr	
	\+ &${\rm b}_r $  &$\>\>1$&$\>\>1$&$-1$   &$-1$  &$\>\>1$   &$[3]$&1/6&$-1/2$&$-1/3$\cr	
	\+ &${\rm b}_g $  &$\>\>1$&$-1$   &$\>\>1$&$-1$  &$\>\>1$   &$[3]$&1/6&$-1/2$&$-1/3$\cr
	\+&|||||||||||||||||||||||||||||||||||||||||\cr
	\+&$q_b(-4/3)$      &$-1$   &$\>\>1$   &$\>\>1$  &$-1$&$-1$ &$[4]$&1/6&$-3/2$&$-4/3$\cr
	\+&$q_r(-4/3)$      &$-1$   &$\>\>1$   &$-1$     &$-1$&$-1$ &$[4]$&1/6&$-3/2$&$-4/3$\cr
	\+&$q_g(-4/3)$      &$-1$   &$-1$      &$\>\>1$  &$-1$&$-1$ &$[4]$&1/6&$-3/2$&$-4/3$\cr
	\+ &&&\cr
	\+&$q_b(5/3)$      &$\>\>1$&$\>\>1$    &$\>\>1$   &$-1$&$-1$&[4]  &1/6&3/2&$\>\>5/3$\cr 
	\+&$q_r(5/3)$      &$\>\>1$&$\>\>1$    &$-1$      &$-1$&$-1$&[4]  &1/6&3/2&$\>\>5/3$\cr 
	\+&$q_g(5/3)$      &$\>\>1$&$-1$       &$\>\>1$   &$-1$&$-1$&[4]   &1/6&3/2&$\>\>5/3$\cr 
	\+&|||||||||||||||||||||||||||||||||||||||||\cr	  	
	\+ &$\nu_e $       &$-1$&$-1$   &$-1$   &$\>\>1$ &$\>\>1$ &[1]  &$-1/2$&1/2&$\>\>0$\cr	
	\+ &$\nu_\mu $     &$-1$&$-1$   &$-1$   &$\>\>1$ &$-1$    &[2]  &$-1/2$&1/2&$\>\>0$\cr	
	\+ &$\nu_\tau $    &$-1$&$-1$   &$-1$   &$-1$    &$\>\>1$ &[3]  &$-1/2$&1/2&$\>\>0$\cr
	\+ &$l(-2)$        &$-1$&$-1$   &$-1$   &$-1$    &$-1$    &[4]  &$-1/2$&$-3/2$&$-2$\cr
	\+ &&&\cr
	\+ &e$^-$       &$\>\>1$&$-1$   &$-1$   &$\>\>1$ &$\>\>1$&$[1]$ &$-1/2$&$-1/2$&$-1$\cr
	\+ &$\mu^-$     &$\>\>1$&$-1$   &$-1$   &$\>\>1$ &$-1   $&$[2]$ &$-1/2$&$-1/2$&$-1$\cr
	\+ &$\tau^-$    &$\>\>1$&$-1$   &$-1$   &$-1$    &$\>\>1$&$[3]$ &$-1/2$&$-1/2$&$-1$\cr 
	\+ &$l(1)$      &$\>\>1$&$-1$   &$-1$   &$-1$    &$-1$   &$[4]$ &$-1/2$&$\>\>3/2$&$\>\>1$\cr	
	\+&|||||||||||||||||||||||||||||||||||||||||\cr}
$$
Corresponding anti-fermions have reversed signs for all quantum numbers, including B and A.
\vfill\eject

\beginsection  Appendix II: $Cl_{7,7}$ generators and quantum numbers of four fermion generations

Clifford Unification [13] is expressed in terms of the $Cl_{7,7}$ algebra, which has 14
mutually anti-commuting generators, denoted $\hat\gamma^p, \{p=1,2..,7\}$ and
$\hat\gamma^q, \{q=a,b,..,g,h\}$. Its physical interpretation is based on the
factorization $Cl_{7,7}= Cl_{3,3}Cl_{2,2}(\alpha) Cl_{2,2}(\beta)$. Fermions are
described seven binary quantum numbers, each of which corresponds to a commuting
element of the algebra. Three quantum numbers are determined by $Cl_{3,3}$,
and two by each of the $Cl_{2,2}$ algebras.

$Cl_{3,3}$ defines the quantum numbers $\rm A =\pm 1,\rm B =\pm 1$ and $\rm C =\pm 1$
which have been shown to describe the spatial properties of fermions [13]. `A' distinguishes the
direction of spins, which have an arbitrary spatial orientation. `B' distinguishes
between fermions (with $\rm B=1$) and anti-fermions (with $\rm B=-1$). `C' determines
fermion intrinsic parities, where $\rm C = 1$ corresponds to the spatial coordinate system
employed in special relativity theory.

Electrons have  $\rm B=C=1$ and charge $\rm Q= -{1\over 2}(B+C) =-1$, while neutrinos
have $\rm B=1,\>C=-1$ and the same charge formula, viz. $\rm Q=-{1\over 2}(B+C) =0$.
This suggests a basis for a general charge formula in the form  $\rm Q=Q_B+Q_C$ where 
$Q_C$ is parity dependent.

$Cl_{2,2}(\alpha)$ has the four anti-commuting generators
$\gamma^4$,  $\gamma^5$,  $\gamma^d$,  $\gamma^e$. These define the commuting
elements $\gamma^{\rm D} =\gamma^{4d}$ and $\gamma^{\rm E}=\gamma^{5e}$. Each 
of these has two eigenvalues, denoted D=$\pm 1$ and E=$\pm 1$,
The quark/lepton distinction is based on the expression D+E-DE= -3 for leptons,
and D+E-DE= 1 for quarks. These values determine the lepton/quark ratio of charge
contributions to $\rm Q_B$. At this stage the fermion charges are given by the
expression
$$
\rm Q= Q_B +Q_C, \>where\> Q_B= {1\over 6}(D+E-DE),\> Q_C= -{1\over 2}C. \eqno({\rm II}.1)
$$

$Cl_{2,2}(\beta)$ has the four anti-commuting generators, namely
$\gamma^6$,  $\gamma^ 7$,  $\gamma^f$,  $\gamma^g$. These define the commuting
elements $\gamma^{\rm F} =\gamma^{6f}$ and $\gamma^{\rm G}=\gamma^{7g}$, each
with $\pm 1$ eigenvalues. The three observable generations have the same fermion
charges and can therefore be associated with F+G-FG = 1 which, taking D,E into account,
gives $\rm Q_C= - {1\over 2}(F+G-FG)C = - {1\over 2}C$. The fourth (unobserved) G(4)
generation has F+G-FG = -3, corresponding to F=G=$-1$, giving $\rm Q_C= {3\over 2}C$,
and the final charge contribution formula for all the elementary fermions
$$
\rm Q= Q_B + Q_C = {1\over 6}(D+E-DEB) - {1\over 2}(F+G-FGB)BC.   \eqno({\rm II}.2)
$$
Factors B have been introduced to make this formula appropriate for anti-fermions
as well as fermions.

\beginsection Acknowledgment

I am grateful to Emeritus Professor Ronald King for pointing out the inconsistency
between quantum number and charge conservation due to the different charges
associated with the C quantum number in G(4) and G(1-3).

\vfill\eject

\beginsection References

\frenchspacing

\item {[1]} Meng, Yue et al (Pandax-4T Collaboration) (2021) Dark Matter Search Results
from the PandaX-4T Commisioning Run, Phys.Rev.Lett.{\bf 127}, 261802

\item {[2]} Bartram, C. et al (ADMX Collaboration) (2021) Search for Axion Dark Matter in
the 3.3 - 4.2 $\mu$eV Mass Range Phys.Rev.Lett.{\bf 127}, 261803

\item {[3]} Belyaev, Alexander, Prestal, Stefan, Rojas-Abbate, Felipe, Zurete (2020)
Probing Dark Matter with Disappearing Tracks at LHC {arXiv:2008.08581v1}

\item {[4]} Foot, R. (2013) Hidden sector dark matter explains the
DAMA, CoGeNT, CRESST-II and CDMS/Si experiments {arXiv: 1209.5602v3}

\item {[5]} G.Adhikari {\it et al} (COSINE-100 Collaboration) (2021) \vskip1pt
Three-year annual modulation search with COSINE-100 {arXiv:2111.08863}

\item {[6]} Krishnak, Aditi, Dantuluri, Aisha and Desai, Shantanu (2020) \vskip1pt
Robust model comparison of DAMA/LIBRA annual modulation {arXiv: 1906.056726v2}

\item {[7]} Bernabei, R., Belli, P., Caracciolo, V., Cerulli, R., Merlo, V., Cappella, F.,
d'Angelo, A. Incicchitti, A., Dai, C.J. Ma, X.H., Sheng, X.D., Montecchia, F. and Ye, Z.P.(2020) 
The dark matter: DAMA/LIBRA and its perspectives {arXiv: 2110.04734v1}

\item {[8]} Carlin, N. (2025) COSINE-100 full dataset challenges the annual modulation
signal of DAMA/LIBRA Sci. Adv.{\bf 11}, eadv6503, {arXiv:2503.19559}

\item {[9]} Thomson, Mark. (2013) Modern Particle Physics (Cambridge University Press)

\item {[10]} Djouadi, Abdelhak (2012) Lenz, Alexander. Sealing the fate of a fourth generation \vskip1pt
{arXiv:1204.1252v2}

\item {[11]} Li, Hsiang-nan (2025) Dispersive determination of the fourth generation lepton masses
J.Phys.G. 52, 025001 {arXiv:2407.07813}

\item {[12]} Li, Hsiang-nan (2025) Dispersive determination of the fourth generation quark masses
Phys.Rev.D {\bf 109}115024 {arXiv:2309.15602}

\item {[13]} Newman, D.J. (2024) Quantum number conservation: a tool in the design and analysis
of high energy experiments J.Phys.G:Nucl.Part.Phys. 51 095002

\item {[14]} Navarro, Julio F., Frenk, Carlos S. and White, Simon, D.M. (1995) \vskip 1pt
The Structure of Dark Matter Halos {arXiv:astro-ph/9508025}

\item{[15]} Roberts, M.G., Braff,L., Garg,A., Profumo,S., and Jeltema,T. (2025) Little Red Dots
from Ultra-Strongly Self-Interacting Dark Matter.{arXiv:2507.03230v2}

\item {[16]} S.Schael et al (The ALEPH, DELPHI, L3, OPAL, SLD Collaborations,
the LEP Electroweak Working Group, and the ALEPH Heavy Flavour Group)
(2006) Phys.Rept.{\bf 427}, 257 {arXiv: hep-ex/0509008}

\item {[17]} Feng, Wei-Xiang, Yu, Hai-Bo and Zhong,Yi-Ming (2021) \vskip1pt
Seeding Supermassive Black Holes with Self-Interacting Dark Matter: \vskip1pt
A Unified Scenario with Baryons. {arXiv:2010.15132}

\item{[18]} Dattathri, Shashank and Sharma, Prateek. (2021) Cosmological evolution \vskip1pt
of gas and supermassive black holes in idealized isolated haloes. {arXiv:2111.04751}

\item{[19]} Busoni, Giorgio (2021) Capture of Dark Matter in Neutron Stars {arXiv:2201.00048}

\item{[20]} Green, Anne M. and Kavanagh, Bradley, J. (2020) Primordial Black Holes
as a dark matter candidate. {arXiv:2007.10722.v3}

\end
\bye